\documentclass[twocolumn,superscriptaddress,prl,aps,amsmath,showpacs]{revtex4}
\usepackage{graphicx}
\usepackage{amsmath,amssymb,amsfonts}

\makeatother
\begin{document}

\title{Synchronization of mutually coupled chaotic lasers in the presence of a shutter}

\author {Ido Kanter}
\affiliation{Department of Physics, Bar-Ilan University,
Ramat-Gan, 52900 Israel}
\author{Noam Gross}
\affiliation{Department of Physics, Bar-Ilan University,
Ramat-Gan, 52900 Israel}
\author{Einat Klein}
\affiliation{Department of Physics, Bar-Ilan University,
Ramat-Gan, 52900 Israel}
\author{Evi Kopelowitz}
\affiliation{Department of Physics, Bar-Ilan University,
Ramat-Gan, 52900 Israel}
\author{Pinhas Yoskovits}
\affiliation{Department of Physics, Bar-Ilan University,
Ramat-Gan, 52900 Israel}
\author{Lev Khaykovich}
\affiliation{Department of Physics, Bar-Ilan University,
Ramat-Gan, 52900 Israel}
\author{Wolfgang Kinzel}
\affiliation{Institut f\"ur Theoretische Physik,
Universit\"at W\"urzburg, Am Hubland 97074 W\"urzburg, Germany}
\author{Michael Rosenbluh}
\affiliation{Department of Physics, Bar-Ilan University,
Ramat-Gan, 52900 Israel}

\begin{abstract}
Two mutually coupled chaotic diode lasers exhibit stable isochronal
synchronization in the presence of self feedback. When the mutual
communication between the lasers is discontinued by a shutter and
the two uncoupled lasers are subject to self-feedback only,  the
desynchronization time is found to scale as $A_d\tau$ where $A_d>1$
and $\tau$ corresponds to the optical distance between the lasers.
Prior to synchronization, when the two lasers are uncorrelated and
the shutter between them is opened, the synchronization time is
found to be much shorter, though still proportional to $\tau$. As a
consequence of these results, the synchronization is not
significantly altered if the shutter is opend/closed faster than the
desynchronization time. Experiments in which the coupling between
two  chaotic-synchronized diode lasers is modulated with an
electro-optic shutter are found to be consistent with the results of
numerical simulations.
\end{abstract}


\maketitle

Chaotic systems are characterized by an irregular motion which is
sensitive to initial conditions and tiny perturbations.
Nevertheless, two chaotic systems can synchronize their irregular
motion when they are coupled \cite{Pikovsky}. When the coupling is
switched off, any tiny perturbation drives the two trajectories
apart. The separation is exponentially fast, and it is described by
the largest Lyapunov exponents of a single system.

In this Letter we show that the trajectory dynamics of coupled
chaotic systems which also poses time-delayed self-feedback, is
different. In a system with self-feedback, which has also been
investigated in the context of secure communication with chaotic
lasers \cite{MCPF}, the time scale for the separation of the
trajectories is found to be much longer than the coupling time. On
the other hand, when the coupling is switched on, resynchronization
occurs on a faster time scale. We investigate this phenomenon
numerically and show first experiments which support our numerical
simulations. The demonstrated difference between de- and
re-synchronization can be used to improve the security of
public-channel communication with chaotic lasers \cite{MCPF}.

Semiconductor (diode) lasers subjected to delayed optical feedback
are known to displays chaotic oscillations. Two coupled
semiconductor lasers exhibit chaos synchronization. Different
coupling setups such as unidirectional or mutual coupling and
variations of the strength of the self and coupling feedback result
in different synchronization states: the two lasers can synchronize
in a leader-laggard or anticipated mode \cite{shore99,locquet01}, as
well as in two different synchronization states; achronal
synchronization in which the lasers assume a fluctuating leading
role, or isochronal synchronization where there is no time delay
between the two lasers' chaotic signals
\cite{IsoPaper,elsasser01,MCPF,exception,Liu1}.

In this Letter we focus on a symmetric setup, the time delay between
the lasers is denoted by $\tau_c$ and the time delay of the
self-feedback is denoted by $\tau_d$. In the event of
$\tau_c=\tau_d=\tau$ and for a wide range of the mutual coupling
strength, $\sigma$, and the strength of the self-feedback, $\kappa$,
the stationary solution is isochronal synchronization
\cite{IsoPaper,MCPF,Gross2006}. The quantity with which we measure
the degree of synchronization between the two lasers is the
time-dependent cross correlation, $\rho$, defined as

\begin{equation}\label{eqrho}
    \rho(\Delta t) = \frac{\sum_i(I_{A}^{i}-<I_{A}^{i }>)\cdot{(I_{B}^{i+\Delta t }-<I_{B}^{i+\Delta t}>})}{\sqrt[]{\sum_i(I_{A}^{i}-<I_{A}^{i }>)^{2}\cdot\sum_i(I_{B}^{i+\Delta t }-<I_{B}^{i+\Delta t}>)^{2}}}
\nonumber\end{equation}

\noindent where $I_{A}$ and $I_{B}$ are the time dependent
intensities of lasers A and B and the summation is over times
indicated by $i$. Isochronal synchronization is defined by the cross
correlation, $\rho$, having a dominant peak at $\Delta t=0$.

We control the mutual coupling between the lasers by a shutter,
located at a distance $c\tau/2$ from each one of the lasers, where
$c$ is the speed of light. When the shutter is open the two lasers
are mutually coupled with strength $\sigma$ and with self-feedback
$\kappa$. When the shutter is closed the self-feedback,
$\kappa_{e}$, is increased to a value of $\sigma+\kappa$ so that the
total feedback in the open/closed states remains a constant.  This
is required so as to prevent a sudden drop in the overall feedback,
which would typically destroy the synchronization immediately
\cite{alhers98}.

The two quantities of interest in this letter are the
desynchronization time, $t_d$, and the resynchronization time,
$t_r$. The desynchronization time is defined as the average required
time for the correlation to decay to $C_d\rho(0)$ where $C_d<1$. The
time is measured from the moment the shutter is closed and $\rho(0)$
is the average correlation in the isochronal phase. The
resynchronization or recovery time is measured after the shutter has
remained closed for a long period and the two chaotic lasers are
uncorrelated, and $\kappa_e=\kappa+\sigma$. The shutter is then
opened and the self-coupling strength is reduced to $\kappa$. The
resynchronization time is defined as the average time required, from
the shutter opening, for $\rho$ to increase from zero to
$C_r\rho(0)$, where $C_r$ is a constant $\le 1$.

To numerically simulate the system we use the Lang-Kobayashi
equations \cite{Kobayashi} that are known to describe a chaotic
diode laser. The dynamics of laser $A$ are given by coupled
differential equations for the optical field, $E$, the time
dependent optical phase, $\Phi$, and the excited state population,
$n$;

\begin{eqnarray}
\frac{dE_{A}}{dt}=\frac{1}{2}G_{N}n_{A}E_{A}(t)+\frac{C_{sp}\gamma[N_{sol}+n
_{A}(t)]}{2E_{A}(t)} ~~~~~~~~~~~~~~~~~~~~~~~~~~~ &  &
\nonumber\\
+ \kappa E_{A}(t-\tau _{d})cos[\omega _{0} \tau + \Phi
_{A}(t)-\Phi_{A} (t-\tau
_{d})]~~~~~~~~~~~~~~~~~\nonumber\\
+ \sigma E_{B}(t-\tau _{c})cos[\omega _{0} \tau _{c}
+\Phi_{A}(t)-\Phi
_{B}(t-\tau _{c})]~~~~~~~~~~~~~~\nonumber\\
\frac{d\Phi_{A}}{dt}=\frac{1}{2}\alpha G_{N}n_{A}
~~~~~~~~~~~~~~~~~~~~~~~~~~~~~~~~~~~~~~~~~~~~~~~~~~~~~~~~~~~~~~ & &
\nonumber\\
-\kappa \frac{E_{A}(t-\tau_{c})}{E_{A}(t)}sin[\omega _{0} \tau +\Phi
_{A}(t)-\Phi_{A}
(t-\tau _{d})] ~~~~~~~~~~~~~~~~~ &  & \nonumber\\
-\sigma \frac{E_{B}(t-\tau _{c})}{E_{A}(t-\tau _{c})}sin[\omega _{0}
\tau _{c} +\Phi_{A}(t)-\Phi _{B}(t-\tau
_{c})]~~~~~~~~~~~~~~~\nonumber\\
\frac{dn_{A}}{dt}=(p-1)J_{th}-\gamma n_{A}(t)-[\Gamma +
G_{N}n_{A}(t)]E_{A}^{2}(t)~~~~~~~~~~~~~~~~~~~ &  &
\nonumber\\
\nonumber\end{eqnarray}

\noindent and likewise for laser B. The values and meaning of the
parameters are those used in Ref. \cite{alhers98,IsoPaper,MCPF}. The
tunable parameters, both in the simulations and in the experiment,
are $\kappa$, $\sigma$ and $p$ the pump parameter which is the ratio
of the actual laser injection current to the threshold current.

Figure 1 displays the calculated desynchronization time as a
function of $\tau$ for $\kappa=\sigma=50ns^{-1}$, $\rho(0)> 0.99$,
p=1.2, and $C_d=0.8$ and $0.9$. Each data point is averaged over
$50$ samples, and $\rho(0)$ is measured by averaging a sliding
window (sliding length is $1$ ns) over a length $\tau$, while the
solid lines are obtained by a linear fit, $t_d = A_d\tau +
constant$. The calculation shows that $t_d$ scales linearly with
$\tau$ with a slope which increase as $C_d$ decreases, and is near
$7$ and $9$ for $C_d=0.9$ and $0.8$, respectively. A similar linear
scaling of the desynchronization time was obtained for all values of
$p$ in the range from $1$ to $1.5$, where for a given $C_d$, $t_d$
decreases with $p$. The simulation assumes an ideal shutter with
instantaneous closing and opening times, and also assumes a
discontinuous decrease of $\sigma$ to zero while $\kappa$ increases
to $\kappa+\sigma$. We find that the linear scaling as well as the
slope  is robust to the following two experimentally necessitated
perturbations: (a) a non-ideal shutter which closes gradually over a
period of $10-20$ nanoseconds; (b) an imperfectly closed shutter,
allowing residual mutual coupling of a few percent of $\sigma$ while
in the closed state.

In the inset of Figure 1, the average $\rho$ as a function of time
for $\tau=50$ $ns$  is presented for a case where the shutter was
closed at t=0. It is clear that the decay of $\rho$ does not consist
of a typical exponential decay. The striking result is that the
correlation coefficient is almost a constant for a long initial
period (first $\sim 250$ $ns$ for the parameters of figure 1) and
then crosses over to an exponential decay for very long times.
Because the event of closing or opening the shutter takes a time
$\tau/2$ to propagate to the lasers it is reasonable to expect that
$t_d$ > $\tau/2$. It is surprising, however, that $t_d$ scales
linearly with $\tau$ with a prefactor which is significantly greater
than $1$.  It is also not obvious from the simulation (in which
$\rho(0) > 0.99$),if such behavior can be observed in an experiment
where $\rho(0)\le 0.9$.

The nearly constant value of $\rho$ after the coupling between the
lasers is terminated calls for a theoretical explanation. Let us
discuss the synchronization for the case $\kappa\sim\sigma$ where
both lasers are driven by an almost identical delayed signal which
is the sum of the self-feedback and the coupling beam. When the
mutual coupling is switched off and replaced with stronger
self-feedback, the system still feels its synchronized state for a
period of length $\tau$, since the system is coupled to its history
delayed by a time $\tau$. The only difference caused by the closing
of the shutter is that the lasers no longer communicate with each
other and each laser is coupled only to its own state. With time, a
small difference in the driving signals develops. This small
difference is amplified, since the system is chaotic, and this
occurs stepwise in time intervals of length $\tau$. For each
interval there is a constant distance between the two trajectories,
which increases for the following interval. Only the envelope of
these steps is described by the largest Lyapunov exponents, but not
the dynamics itself. Hence desynchronization is very slow, and its
time scale should be proportional to $\tau$. On the other hand, when
the exchanged beam is switched on again, both lasers receive
immediately an identical feedback signal (for $\kappa=\sigma$) and
synchronize very fast, independent of $\tau$. We thus expect that
the desynchronization time is insensitive to the values of $\kappa$
and $\sigma$ and should scale with $\tau$. In contrast the
resynchronization time should be very sensitive to the difference,
$\kappa-\sigma$.

\begin{figure}
\vspace{-0.5 cm} {\centering
\resizebox*{0.4\textwidth}{0.22\textheight}
{{\includegraphics[angle=270]{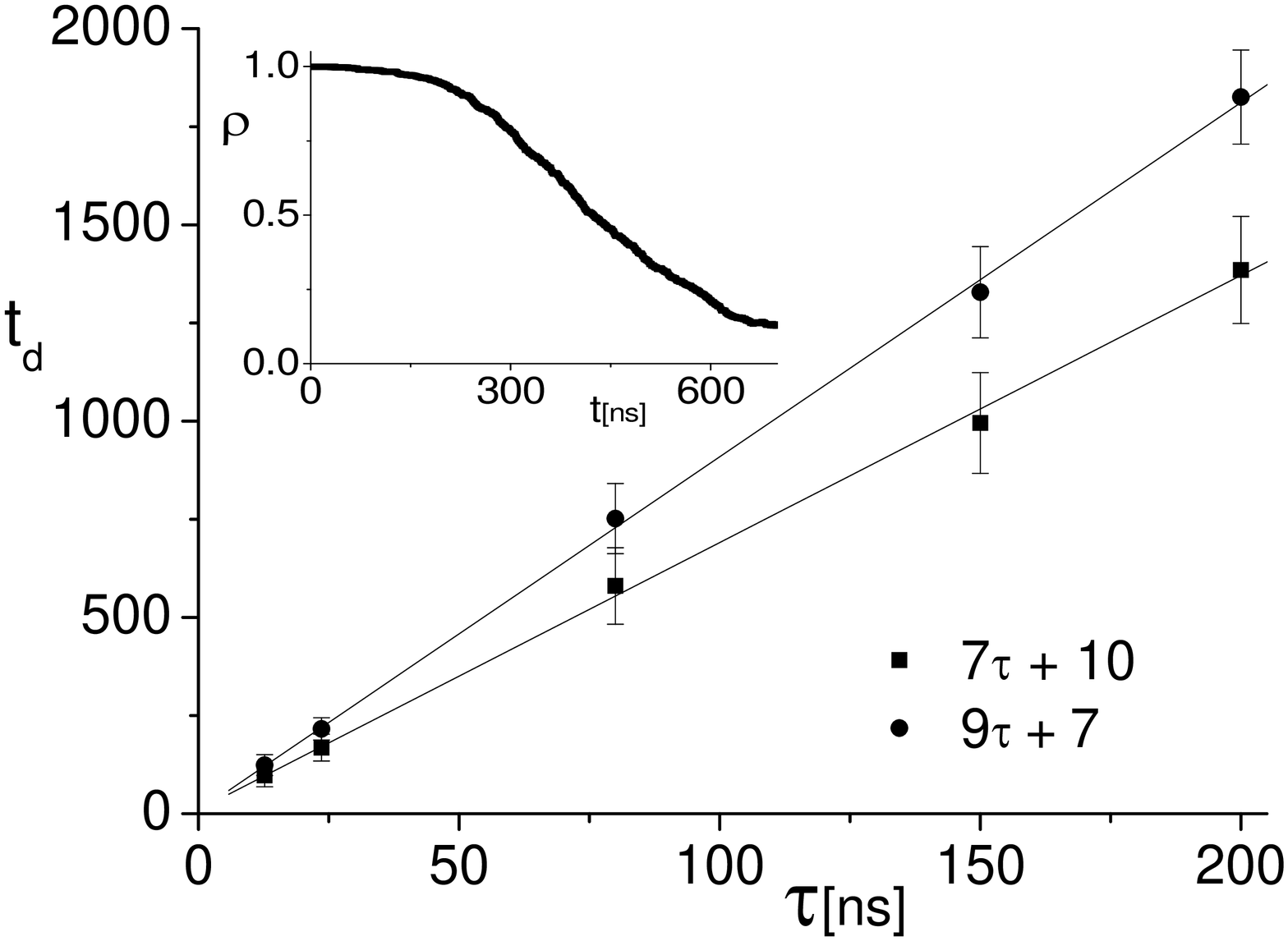}}}
\par}
\caption{$t_d$ as a function of $\tau$ for p=1.2, $\kappa=\sigma=50
ns^{-1}$ and $C_d=0.9$ ($\blacksquare$) and $C_d=0.8$ ($\bullet$).
The lines are a result of a linear fit and the error bars obtained
from 50 independent calculations for each point. Inset: The average
$\rho$ as a function of time for the same parameters as in the main
figure and $\tau=50ns^{-1}$.}
\end{figure}

The experimental setup, which confirms many of these numerical
perdictions, is shown schematically in Figure 2. We use two
semiconductor lasers emitting at $660$ $nm$ and operated close to
their threshold currents. The temperature of each laser is
stabilized to better than $0.01K$. The lasers are subjected to
similar optical feedback and are mutually coupled by injecting a
fraction of each one's output power to the other. A fast
electro-optic modulator, with measured closing/opening time of 15
$ns$, is introduced in the middle of the coupling optical path to
enable closing and reopening of the mutual coupling.  The optical
setup is designed so as to compensate the sudden drop in the overall
feedback power when the shutter is closed and the mutual coupling
feedback drops to near zero. We thus use the shutter as a
polarization beam splitter which divides the output power of the
laser into two parts: one used for the self-feedback and the other
for the mutual coupling channels. The opening and closing of the
shutter merely changes the ratio of powers in the channels, but
maintains the overall feedback power at a constant level. Without
this precaution the sudden drop in feedback power would destroy
synchronization immediately. This setup, however, does not prevent
the change in phase of the laser field when the shutter changes its
state. This residual effect decreases the level of synchronization
by a small amount (as can be seen in Figure 3). The shutter also
does not close hermetically and the leakage power to the mutual
coupling channel in the closed state is measured to be $\sim$ 7\% of
the shutter open value.

\begin{figure}
{\centering \resizebox*{0.42\textwidth}{0.22\textheight}
{{\includegraphics[angle=270]{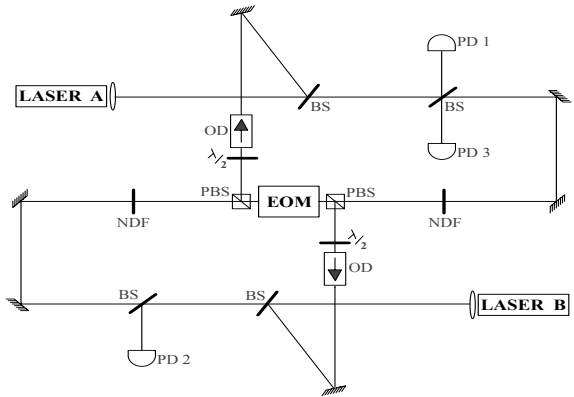}}}
\par}
\caption{\label{schema} Schematic of the two mutually coupled
lasers. EOM - Electro-optic modulator; BS - Beam Splitter; PBS -
Polarizing Beam Splitter; OD - Optical Isolator; NDF - Neutral
Density Filter; PD - Photodetector.}
\end{figure}

The feedback strength of each laser is adjusted using a $\lambda/$2
wave plate and an optical diode (see Figure \ref{schema}) and is set
to a value which leads to a reduction of about 5$\%$ in the solitary
laser's threshold current \cite{Gross2006}. The lengths of the
self-feedback and coupling optical paths are set to be equal to
obtain stable isochronal synchronization \cite{MCPF}. Two sets of
measurements are reported here, corresponding to two self-feedback
optical paths with $\tau$ = $12.7$ $ns$  and $\tau$ = $23.6$ $ns$.

Two fast photodetectors (response time $<$ 500 $ps$) are used to
monitor the laser intensities which are simultaneously recorded with
a digital oscilloscope (500 MHz ,1 GS/sec). The correlation
coefficient, $\rho$, is calculated by dividing the intensity traces
into 10 $ns$ segments (each segment containing 10 points) and $\rho$
is calculated between matching segments and then averaged.

The measured correlation coefficient, $\rho$, is shown in Figure 3,
in which the shutter is closed at t=0. The coupling power decays to
its closed level in about 15 $ns$, limited by the speed of the
shutter. The observed decay time is only slightly shorter than the
decay time obtained in simulations for the same $\tau$, and as in
the simulations, $\rho$ initially maintains a high and nearly
constant value for $50-100$ $ns$, which is much longer than $\tau$.
The four data curves shown in the figure correspond to different
experimental parameters, indicated by the value of $p$ and to the
two different values of $\tau$.

\begin{figure}
\vspace{-0.4 cm}
{\centering\resizebox*{0.45\textwidth}{0.22\textheight}
{{\includegraphics[angle=270]{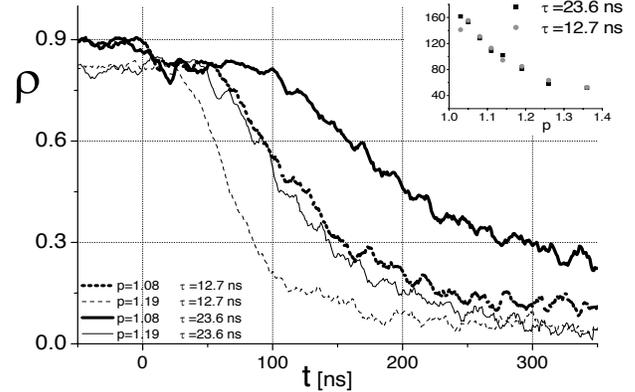}}}
\par}
\vspace{-0.3 cm} \caption{Experimental results for
$\rho$ as function of $t$ for $p=1.08,~1.19$ and $\tau=12.7,~23.6$
$ns$ . Each data point was averaged over 500 measurements. Inset:
Data collapse, $1.7\cdot t_d$ ($t_d$) measured in $ns$ as a function
of $p$ for $\tau=12.7$ $ns$ ($23.6$ $ns$).}
\label{recovery}
\end{figure}

Figure 3 shows that $t_d$ increases with $\tau$ and the inset of
Figure 3, depicting  $t_d$ as a function $p$, indicates that $t_d$
is a decreasing function of $p$. The inset of Figure 3 also
demonstrates that the different decay curves all collapse to a
single decay curve, independent of $p$, when scaled by a factor of
$1.7$ which is very close to the ratio of $\tau_1 / \tau_2=23.6/12.7
\sim 1.86$. The numerical simulations for larger $\tau$  also
exhibit such data collapse when scaled by $\sim \tau_1/\tau_2$
resulting in scaled decay curves which are independent of
$p$\cite{prep}. This result and the linear scaling of $t_d$ for a
given $p$ indicate
\begin{equation}
t_d(\tau,p) \propto \tau g(p)
\end{equation}
\noindent where $g(p)$ is a function characteristic of the specific
diode laser used. For small $\tau$ finite size effects are expected
as a result of the positive constant in the linear scaling shown in
Figure 1. Indeed, for $\tau=12.7$ and $23.6$ $ns$, the numerical
results indicate that the average ratio $t_d(\tau_1,p)/t_d(\tau_2,p)
\sim 1.75$, which is in surprisingly good agreement with the
experimental result of $1.7$.

The simulations also indicate that a transition from the low
frequency fluctuation (LFF) regime to the fully developed coherent
collapse regime \cite{FCDC} occurs at $p \sim 1.35$, which is close
to the experimental value of $p$ (inset of Figure 3) where $t_d$
becomes almost independent of $p$\cite{prep}. Though the decay time
obtained from the simulations is longer that the decay time observed
in the experiment, this is not surprising, since in the simulations
the systems are initially correlated to a very high level
($\rho(0)>0.99$) while in the experiments the initial correlation is
$\rho(0)\le 0.9$

We now turn to examine the scaling of the resynchronization time or
the recovery time as a function of $\tau$.  In the simulations we
start with two uncoupled systems with self-feedback
$\kappa_e=\kappa+\sigma$. When the shutter is opened at $t=0$,
$\kappa_e$ is reduced discontinuously to $\kappa$ and the mutual
coupling is set to $\sigma$. For all examined cases, with $\kappa
\ne \sigma$, our calculations indicate that the resynchronization
time also scales linearly with $\tau$.  This scaling is exemplified
in Figure 4 for $C_r=0.9$, $\kappa=60 ns^{-1}$ and $\sigma=40
ns^{-1}$.

\begin{figure}
\center{\includegraphics[angle=270,width=3.5in]{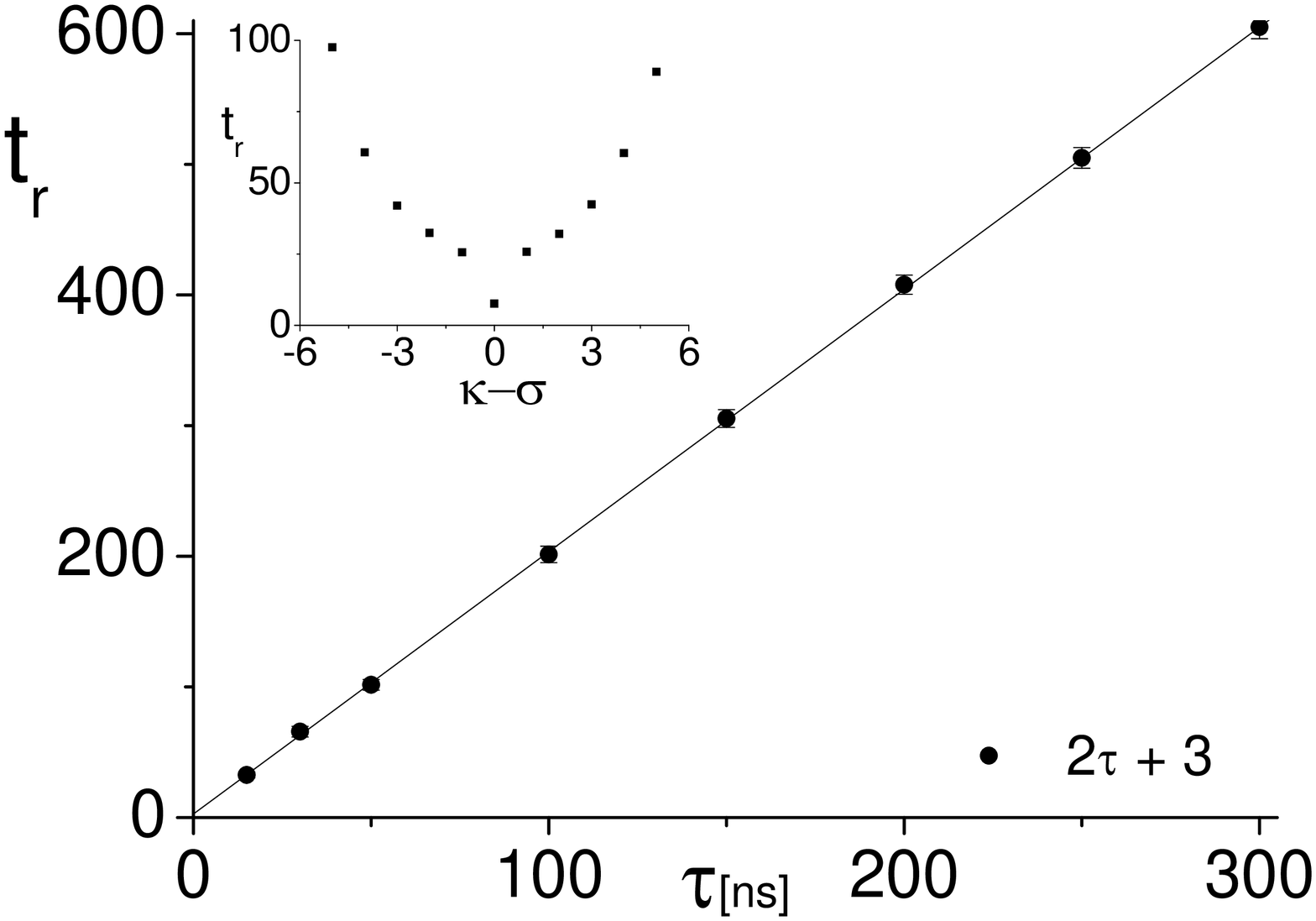}}
\caption{$t_r$ as a function of
$\tau$ for $p=1.2$, $\kappa=60 ns^{-1}$ and $\sigma=40 ns^{-1}$.
Inset: $t_r$ as a function of $\kappa-\sigma$ for $\tau=15$ $ns$. }
\label{recovery1}
\end{figure}

\begin{figure}
\center{\includegraphics[width=3.5in]{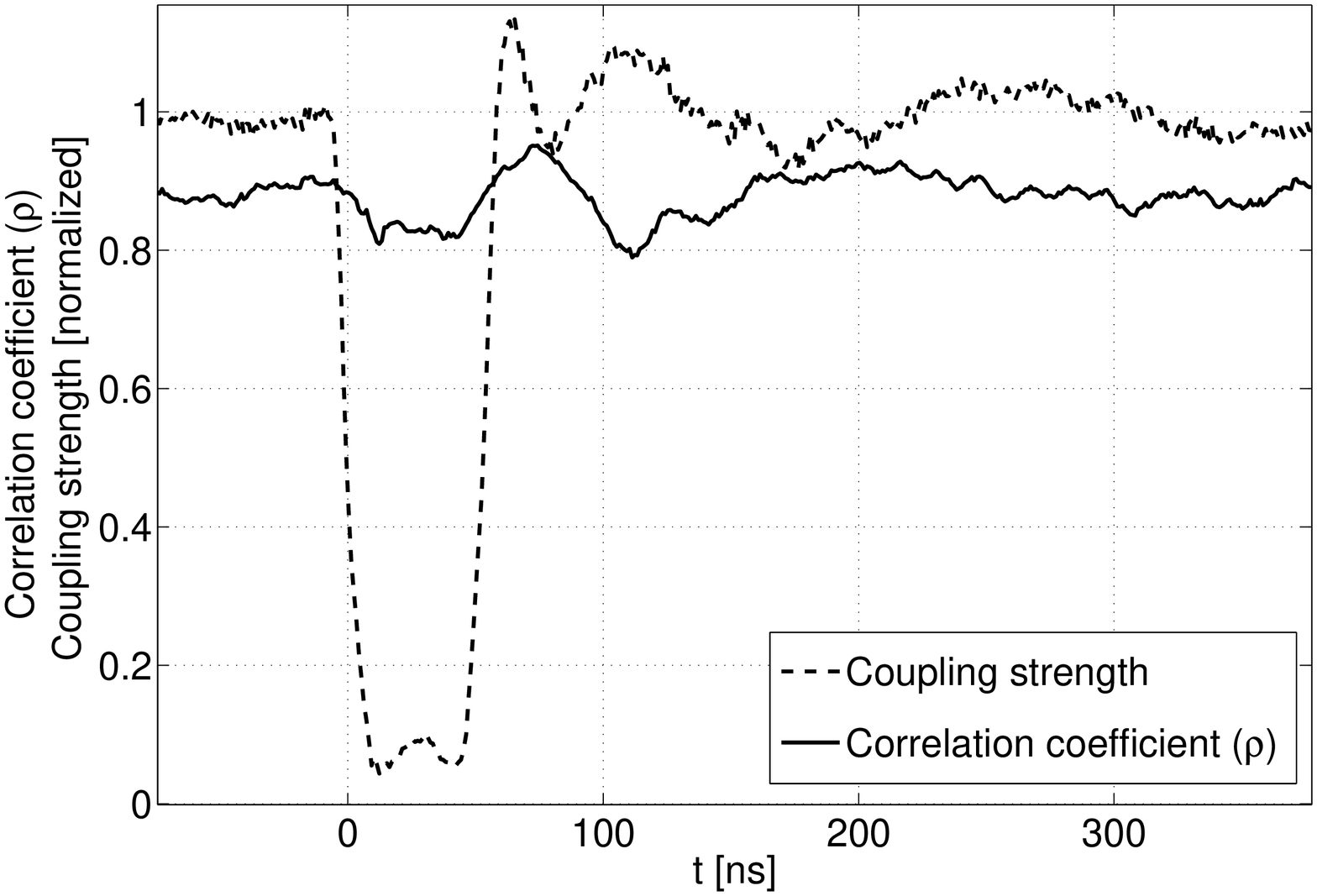}}
\caption{Normalized coupling
strength, $\sigma$, (dashed line) and correlation strength, $\rho$,
(black line) as a function of time.}
\label{recovery2}
\end{figure}

The inset of Figure 4 displays the resynchronization time as a
function of $\kappa-\sigma$ for a given $\tau$. It appears that this
difference, rather than the coupling strength itself, is what
controls the resynchronization time, and $t_r$ scales almost
linearly and symmetrically with $\kappa-\sigma$. For
$\kappa=\sigma$, $t_r$ is very fast and simulations indicate that it
is independent of $\tau$ (limited by the fixed size of the sliding
window) as expected. In all examined cases, the prefactor of the
linear scaling of the resynchronization time was found to be $\ll
A_d$, indicating $t_r < t_d$. We also observed similar behavior, i.
e. $t_r<t_d$, in the experiment, though quantitative determination
of the experimental resynchronization time is complicated by ringing
in the high voltage electronics used to turn on the modulator and by
the fact that the modulator response time is as long as 15 $ns$,
which is comparable to $t_r$.

Although experimentally we cannot accurately measure the
resynchronization time, we show, in Figure 5, a demonstration of the
persistence of the synchronization between the lasers upon repeated
closing-opening operations of the shutter. Shown is the typical
behavior of $\rho$ while the shutter is closed for $\sim40$ $ns$ and
then reopened. The other parameters of the experiment are
$\tau=12.6$ $ns$ and $p \sim 1.08$. The cross correlation
coefficient $\rho$ is not affected, by the closing/reopening of the
shutter and the changing of $\kappa_e$ and $\sigma$, since $t_d
>40$ $ns$.

The results reported above for re/de-synchronization times, which
were also obtained recently for chaotic maps \cite{prep},
demonstrate the possibility of establishing a reliable chaos based
communication channel even while the communication between the
lasers is interrupted   by relatively long intervals. We expect that
these effects will play an important role in advanced secure
communications using mutually chaotic lasers.

\end{document}